\documentclass[]{revtex4-2}
\usepackage[utf8]{inputenc}
\usepackage{graphicx}
\usepackage{epstopdf}
\usepackage{amssymb}
\usepackage{amsmath}
\usepackage{hyperref}
\usepackage{mathrsfs}
\usepackage{caption}
\usepackage{subcaption}
\usepackage{color,soul}
\usepackage{booktabs}
\usepackage[dvipsnames]{xcolor}
\usepackage[toc,page]{appendix}
\usepackage{tikz}
\usepackage{framed}
  \definecolor{shadecolor}{named}{Yellow}
\usepackage{dcolumn}% Align table columns on decimal point
\usepackage{bm}% bold math
\usepackage{framed,graphicx,xcolor}
\usepackage{ragged2e}

\newcommand{\tens}[1]{%
  \mathbin{\mathop{\otimes}}}%

\begin{document}
\title{A geometrically inspired constitutive framework for damage and intrinsic self-healing of elastomers}
 \author{Sanhita~Das}%
\email{sanhitadas@iitj.ac.in}
\affiliation{%
 Department of Civil and Infrastructure Engineering,
Indian Institute of Technology Jodhpur, Jodhpur 
}
\author{Nivedita~Kumari}
\affiliation{Department of Civil and Infrastructure Engineering,
Indian Institute of Technology Jodhpur, Jodhpur }
%\date{September 2019}

\begin{abstract}
Autonomic interfacial self-healing in elastomers enables their reprocessing and recycling, making them valuable for applications such as ballistic resistance, surface coatings, adhesives, and biomedical materials. This article prescribes a geometry-based damage-healing theory for autonomic healing in elastomers, built on a framework where damage induces an incompatibility in the Euclidean material manifold, transforming it into a Riemannian manifold. Healing restores the Euclidean state through a reversing damage variable or an evolving healing variable. The reversing damage variable models the rebonding mechanism while the healing variable accounts for healing by chain diffusion and entanglement. The model also predicts healing where rebonding is preceded by chain diffusion. The microforce balance governs the evolution of the damage and healing variables, capturing rate-dependent damage and intrinsic temperature-independent healing. The model is validated through numerical simulations, including one-dimensional and two-dimensional simulations, demonstrating accurate predictions of coupling between damage and healing and post-healing mechanical response. The model also predicts the recovery of fracture toughness with healing time in supramolecular elastomers, aligning with experimental data. With minimal parameters, the model is versatile and can easily be used for material design and structural analysis, surpassing existing models in simplicity and predictive capability.
\end{abstract}

\maketitle

\section{Introduction}

Self-healing behaviour is introduced into elastomers so that they can be reprocessed and recycled. The autonomic interfacial self-healing observed in a wide range of elastomers is particularly intriguing to both industry and the scientific community. Elastomers with this capability are used for ballistic resistance, tires, surface coatings, adhesives, and biomedical applications such as artificial skins, tissue engineering, etc. \cite{zhao2016self}. Autonomic or stimulus-independent interfacial self-healing is a type of intrinsic self-healing that starts on its own when broken pieces touch each other without any outside help. Understanding and predicting this phenomenon is necessary for the development of efficient self-healing systems, mimicking tissue healing and having proper control of the healing parameters. While experiments are crucial, they are prohibitively expensive, fail if there are multiple healing mechanisms, and have limited capabilities for estimating healing efficiencies, determining the mobility of polymeric chains, etc. \cite{grande2015interfacial}. To help scientists figure out how to heal things and other factors that might affect the connection between damage and healing, damage-healing constitutive models are used along with experiments. They impart information on the recovery of fracture and elastic properties with changes in healing time and other environmental conditions. This, in turn, can aid in the design and study of new self-healing concepts, as well as the development of self-healing elastomers that exhibit more attractive mechanical behaviour under various stress conditions, thereby enabling engineers to predict performance and longevity more accurately. By integrating these models into computational frameworks, researchers can simulate real-world scenarios and optimise the material design process further. \cite{rhaman2011self}. Furthermore, damage-healing models can give useful information with few but measurable macroscopic properties. They can be used in computer models to look at how structures made with these materials fail and how easily they can be recycled. 

Chain diffusion and entanglement \cite{yamaguchi2009interdiffusion} or the rebonding of broken chains \cite{liu2013self,cao2021mechanical,sani2022intrinsic,bekas2016self} are the main processes that make autonomic interfacial healing happen. Rebonding is a chemical process, while chain diffusion is a physical process. It can happen through covalent interactions, such as the Diels–Alder reaction products, disulphide and diselenide bridging, or non-covalent interactions, such as hydrogen bonds, ionic or hydrophobic interactions, and so on. There may be different rebonding mechanisms depending on the type of reversible bonds available. Supramolecular elastomers, which are used in surface coatings and ballistic resistance \cite{cordier2008self, grande2015interfacial}, heal by rebonding. The dynamic bonds in these materials are hydrogen bonds that can be broken reversibly. Rebonding is also observed to be preceded by the diffusion of chains in some elastomers \cite{amaral2017stimuli}. 

Irrespective of the type of mechanism, autonomic healing has strong temporal and spatial dependence \cite{utrera2020evolution}. As the healing time increases \cite{cordier2008self, grande2015interfacial}, chains that are diffusing across the interface move farther, and there are more re-association events. The extent of healing is also dependent on the time gap between the damage and the healing in the case of re-bonding since, with time, the number of potential connections decreases. Additionally, healing depends on the localisation and scale of the damage, whether it is a superficial, wide scratch or a deep crack. Wider damage takes a longer time to heal, according to \cite{amaral2017stimuli}, especially when the mechanism is chain diffusion, as chains take a longer time to travel. The extent of healing has a direct influence on the post-healed damage response of these materials. In general, an increase in healing time increases the stiffness and the strain of fracture of the healed material \cite{darabi2012continuum}, but that may vary with the mechanism. For instance, in \cite{cordier2008self}, the healed elastic response of the supramolecular elastomer is only dependent on the waiting time before healing commences and is completely independent of the healing time. The strain on the fracture, however, increases with healing time. 

Experiments have been, so far, the primary means of understanding healing mechanisms, designing newer self-healing polymers, and predicting healing in actual engineering systems. Only a few attempts have been made to model the phenomena. Submacroscopic models, particularly molecular dynamics simulations of interfacial healing and surface welding \cite{zheng2021molecular, sun2020molecular, chen2020molecular,xu2020structural}, provide only atomistic scale insights into rebonding dynamics, timescales, influence cross-linking specific to disulphide bonds or DA reactions, relaxation timescales, etc. However, they cannot predict responses to actual fracture and healing phenomena in engineering problems. Coupling these models with actual structural analysis or finite element simulations of failure analysis is infeasible, as the resulting multiscale models require a huge computational effort. These models don't offer any understanding of post-healed stiffness and fracture properties either, which could be readily used in solving these larger-scale engineering problems. However, there are microscopic models by \cite{wool1981theory, wool1985properties} that determine some scaling laws for post-healing fracture toughness and stiffness, but they are limited to healing mechanisms involving diffusion and entanglement only, with no rebonding. The scaling laws, however, are physical, as established by the experiments of \cite{grande2015interfacial}. Also, they have found use in subsequent theoretical studies on the dependence of healing time on post-healed mechanical response, but not in real-life engineering damage-healing problems. Mesoscale and multiscale models that bridge submacroscopic scales may be more efficient than molecular dynamics simulations but in no way superior to continuum scale models \cite{shojaei2020continuum} in terms of simplicity, computational efficiency, and the number of material parameters. Most continuum scale models predict damage and healing within a single framework, considering that the underlying mechanisms are strongly coupled. Such coupled damage-healing prescriptions find maximum use in the analysis of real-life engineering structures as well as in material design.

Although researchers have sufficiently explored the microscopic and molecular dynamics modelling of the healing process, there are very few continuum scale models available. The elastoplastic mesoscopically informed continuum scale models of \cite{yu2018mechanics, yu2020mechanics} predict interfacial healing in general dynamic polymeric networks, taking into account diffusion, entanglement, and rebonding mechanisms within a single framework. Being comprehensive, the model is complex, requiring a large number of material parameters that can be determined either through mesoscopic experimental investigations or rigorous and time-consuming curve fitting similar to \cite{das2019constitutive}. This makes them unsuitable for both material design and implementation when analysing actual damage-healing experiments.

The continuum Damage and Healing Mechanics (CDHM) \cite{barbero2005continuum, miao2015constitutive} framework has been widely used to prescribe damage-healing models for a large range of materials. The CDHM framework employs evolving internal variables to measure the damage and healing processes. Typically, the damage variable is defined as the ratio of the damaged to the undamaged area \cite{lemaitre1985continuous, murakami1988mechanical}, while the healing variable is the fraction of the damaged area that has been recovered. The damage variable, scaled by the healing variable, defines the effective damage variable. Instead of area or volume, the variables may also be defined in terms of mechanical or electrical properties of the damaged and the undamaged material. Multiple healing and damage mechanisms can be defined by separate internal variables with independent evolution equations. The framework itself does not pose any restrictions on the evolution equations. For restrictions of thermodynamic consistency, the rational thermodynamics framework \cite{coleman1967thermodynamics} can be invoked where constitutive forms can also be determined from the Coleman-Noll procedure. In addition to the definition of the internal variables, CDHM invokes the concept of scaling the stress by a factor dependent on the effective damage variable. One of the biggest advantages of the rational thermodynamics-informed CDHM framework is that it can be used to formulate local and non-local damage-healing models of any material that can account for multiple driving mechanisms, both extrinsic and intrinsic. \cite{oucif2018modeling,voyiadjis2011thermodynamic,voyiadjis2012generalized,voyiadjis2012continuum,voyiadjis2012theory,darabi2012continuum,al2010micro,mergheim2013phenomenological,sanz2019numerical}. prescribe damage-healing models for polymers, polymer blends, polymers reinforced with shape memory polymers, etc. They predict damage and healing with sufficient accuracy but are parameter-intensive and complex, requiring chain-level material parameters. While the evolution equations in some of these models are phenomenological with no relevance to the actual underlying physics, there are some models with no length-scale parameters to model diffusive processes.

Local damage-healing models fail to predict diffused damage and the influence of the damage width on the rate of healing, which is extremely crucial in the context of healing of scratches and superficial damage \cite{amaral2017stimuli}. Also, for numerical stability of algorithms, damage evolution equations must have non-local terms \cite{das2021geometrically}. \cite{kumar2020phase, kumar2020nucleation} prescribe phase-field-inspired damage-healing models for microcrack-driven fracture in PDMS. The ordinary phase-field model with quadratic shape functions \cite{das2021geometrically}, is not appropriate for elastomers that undergo long stretches prior to brutal damage accurately. Moreover, the mathematically motivated surface energy prescription fails to reflect the timescales associated with the healing process. To alleviate these typical drawbacks of a regularised continuum gradient-based damage model, a geometrically inspired damage model is described in \cite{das2021geometrically}. Damage in the form of polymeric chain scission changes the manifold from Euclidean to Riemannian, which leads to an incompatibility that can be measured by a Ricci curvature. In line with \cite{lemaitre1985continuous, murakami1988mechanical}, a scaling of the metric is proposed such that the damage variable may physically represent the ratio of damaged and undamaged areas. We use the Ricci curvature to define the resistive part of the free energy density, the variation of which determines the evolution of the damage variable. This model is numerically stable and provides a much more accurate prediction for brutal damage compared to a quadratic phase-field damage model. 

In this article, we propose a geometrically inspired non-local, coupled damage-healing model that predicts damage and intrinsic self-healing in elastomers. It is built on the geometric framework \cite{das2021geometrically} and borrows concepts from Continuum Damage Healing Mechanics (CDHM) \cite{barbero2005continuum}. The current theory, in addition to predicting rate-dependent damage, reflects the recovery of the damaged area through an independently evolving healing variable and a reversing damage variable. The healing variable represents the area recovered due to entanglement and diffusion of chains, while the reversing damage variable, which has its own timescale, models healing due to bond reversal. The Ricci curvature defines the resistive component in the free energy expression, which governs the evolution of both the healing and damage variables. The microforce balance laws \cite{gurtin2010mechanics} yield strongly coupled, non-local evolution equations for the damage and healing variables from which models for rate-dependent damage and healing models for chain diffusion and rebonding may be derived. The rational thermodynamic framework of \cite{coleman1967thermodynamics} is used to determine the constitutive restrictions on the stresses and microforces.

The damage and the healing models are implemented in single material points, one-dimensional and two-dimensional simulations of single-cycle or multicycle fracture and healing experiments to investigate the influence of healing time, initial damage width, and healing timescales on the post-healing behaviour and fracture properties of the elastomer. Further, the model is used to investigate the recovery of a supramolecular elastomer's fracture toughness with healing time from the experiments of \cite{cordier2008self}.

The first section of the article \ref{Reimannian Geometry and Kinematics} describes the kinematics, defining the different configurations. Next is the \ref{freeenergy} section, which finds the damaged-healed material's free energy. This is followed by the development of the macroforce and microforce balances in \ref{Macroforce and Microforce balance}. The constitutive prescriptions for the macro and micro stresses are derived in \ref{Constitutive theory}. This is followed by the section \ref{miscellaneoussimulations}, which describes the numerical studies carried out using the model. Finally, the article is concluded in \ref{Conclusion}. The article also includes an appendix \ref{PuredamageSelf-Healing} where a relationship between Griffith's fracture toughness $G_c$ and the resistive modulus $M$ is derived for a pure damage case.

\section{Kinematics of the Damaged/healed Undeformed and Deformed Riemannian Manifold}\label{Reimannian Geometry and Kinematics}

\begin{figure}
\centering
\includegraphics[width=0.8\textwidth]{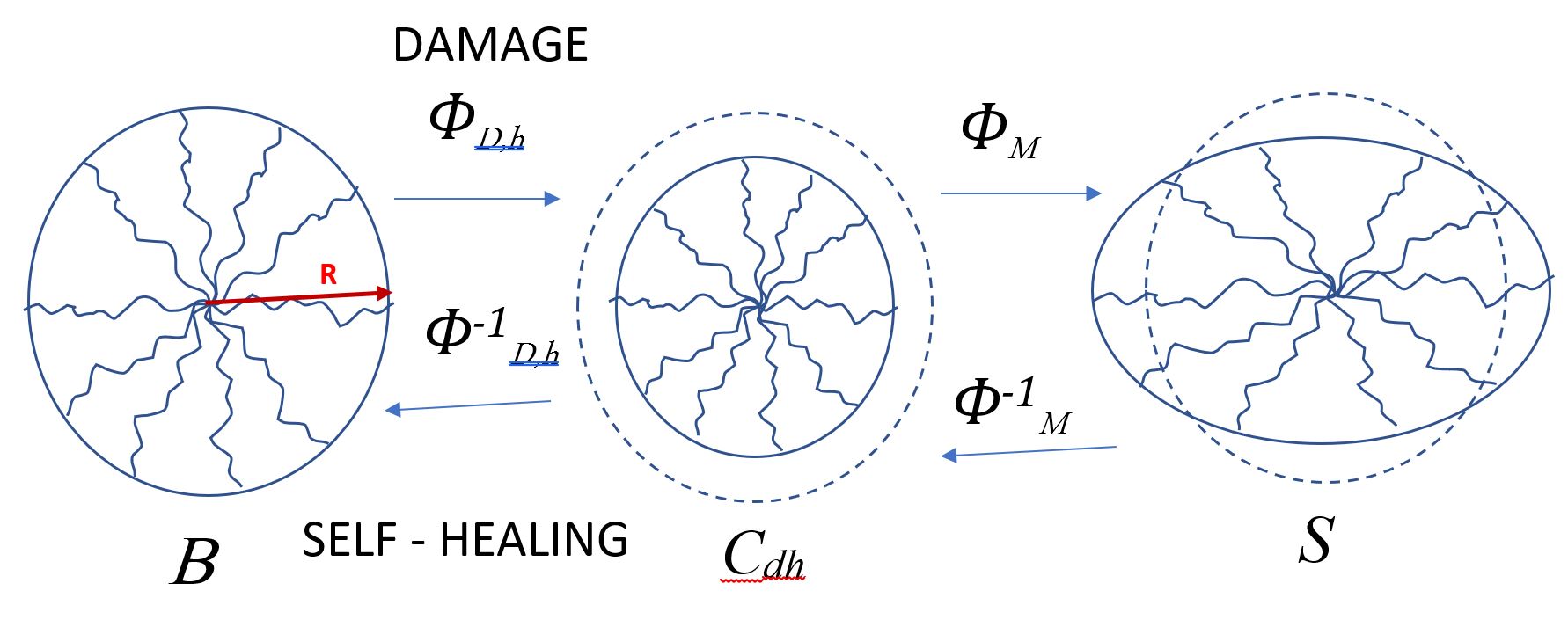}\centering
\caption{\label{Configurations} Schematic figure of the three configurations and the deformation maps}
\end{figure}

\cite{das2021geometrically} prescribes a geometrically inspired damage model for elastomers that shows better convergence, which is a measure of the reduction in area that can effectively contribute to the material's stiffness. This change in area is non-conservative and may be interpreted to change the Euclidean manifold to a Riemannian manifold, as schematically shown in \ref{Configurations}. We define three material configurations the referential or the undamaged, undeformed configuration $\mathcal{B}$, the intermediate fictitious effective damaged-self-healed and undeformed configuration $\mathcal{C}_{dh}$ and the deformed damage configuration $\mathcal{S}$. As in case of \cite{das2021geometrically}, the configuration  $\mathcal{B}$ is Euclidean, while the configurations $\mathcal{C}_{dh}$ and $\mathcal{S}$ are Riemannian manifolds. The system's kinematics reflects the physical change in the area primarily through a simple metric scaling. The scaling is a function of the ratio of the effective area to the undamaged area $D_h$ which is a function of the damage variable $D$ and the healing variable $h$. $D$ as in \cite{das2021geometrically} denotes the ratio of the number of broken polymeric chains to the intact chains in a spherical RVE. $h$, on the other hand, is the ratio of the number of recovered bonds to the number of severed bonds. It may also measure the healing process by other mechanisms, such as the increase in the number of entanglements or cross-links, which can add stiffness to the material. The effective damaged area is used to scale the metric of the undeformed, undamaged configuration, which is Euclidean, to yield the metric of the deformed, damaged manifold. The incompatibility associated with the Riemannian manifold arises due to the damage and healing. The healing contributes to the incompatibility only if the process entails some permanent change to the configuration other than the reversal of a bond-breaking process.  

$\phi$ the isomorphic map from the reference to the deformed, effective configuration is a composition of two isomorphic maps $\phi_{dh}$ and $\phi_M$ 
 defined by,
 \begin{equation}\label{Differ_dam_1}
    \phi_{dh} : \mathcal{B} \rightarrow \mathcal{C}_{dh}
\end{equation}
\begin{equation}\label{Differ_dam_2}
    \phi_M : \mathcal{C}_{dh} \rightarrow \mathcal{S}
\end{equation}

We also define the tangent maps or the deformation gradients $\mathbf{F_{dh}}$ and $\mathbf{F_M}$  respectively. 

\begin{equation}\label{Differ_dam_3}
\mathbf{F}=\mathbf{F_M}\mathbf{F_{dh}}
\end{equation}

Additionally, we assume $\phi_{dh}$ to be an identity map; thus, the associated deformation gradient $\mathbf{F}_{dh}$ mapping the respective tangent spaces is identity. Thus the equation \eqref{Differ_dam_3} simplifies to 
\begin{equation}\label{Differ_dam_4}
\mathbf{F}=\mathbf{F_M}
\end{equation}

The corresponding inverse map $\phi^{-1}$, which maps the damaged, deformed configuration to the undamaged, undeformed configuration, is again a composition of $\phi_{dh}^{-1}$ and $\phi_M^{-1}$ with the following definition
 \begin{equation}\label{Differ_dam_1}
    \phi_{dh}^{-1} : \mathcal{C}_{dh} \rightarrow \mathcal{B} 
\end{equation}
\begin{equation}\label{Differ_dam_2}
    \phi_M^{-1} : \mathcal{S} \rightarrow \mathcal{C}_{dh}
\end{equation}

Associated with the manifolds are the respective metric tensors $\mathbf{G}$, $\mathbf{G_{dh}}$ and $\mathbf{g_{dh}}$. The metric tensor associated with undamaged Euclidean configuration is $\mathbf{G}$ those with undeformed and deformed Riemannian manifold is denoted by $\mathbf{G_{dh}}$ and $\mathbf{g_{dh}}$ respectively,
The undeformed and deformed Riemannian metrics can be expressed as functions of the undeformed and deformed damaged configurations as follows,

\begin{equation}\label{Differ_dam_5}
    \mathbf{G_{dh}} = f(D,h) \mathbf{G}
\end{equation} 
\begin{equation}\label{Differ_dam_6}
    \mathbf{g_{dh}} = f(D,h) \mathbf{g}
\end{equation}

where $\mathbf{G}$ and $\mathbf{g}$ are Euclidean metrics of the undamaged, undeformed and undamaged, deformed configurations. $f(D,h)$ is a scaling factor, a measure of the ratio of the effective damaged area to the area before the damage. In line with \cite{darabi2012continuum}, the scaling factor is chosen as $f(D,h) = (1-D(1-h))$.
Pulled back to the undeformed undamaged configuration, one can write the right Cauchy Green tensor associated with the deformation as

\begin{equation}\label{Differ_dam_7}
    \mathbf{C_{dh}} = f(D,h) \mathbf{F}^{T} \mathbf{g} \mathbf{F}
\end{equation}

where we make use of \eqref{Differ_dam_4}. 

Using Miehe's microsphere model \citet{miehe2004micro}, we calculate the homogenised stretches, area and volume measures in the various configurations that appear in the free energy. In the effective configuration, the homogenised stretch is given as,

\begin{equation}\label{Differ_dam_9}
    \Lambda_{dh0}^2 = (1-D(1-h))
\end{equation}

While in the deformed configuration, the homogenised stretch is given by,

\begin{equation}\label{Differ_dam_11}
    \Lambda_{dh}^2 = \frac{(1-D(1-h))}{3} tr(\mathbf{C})
\end{equation}

where $\mathbf{C} = \mathbf{F}^{T} \mathbf{g} \mathbf{F}$

The infinitesimal area measures for the deformed configuration $dA_{dh}$ are given by,

\begin{equation}\label{Differ_dam_13}
    dA_{dh} =  (1-D(1-h)) dA_0
\end{equation}

The volume measure in the effective configuration,

\begin{equation}\label{Differ_dam_15}
    J_{dh0} = (1-D(1-h))^{3/2}
\end{equation}

and for the deformed damaged manifold, it is given by, 

\begin{equation}\label{Differ_dam_16}
    J_{dh} = (1-D(1-h))^{3/2} \sqrt{det(\mathbf{F}^T\mathbf{F})}
\end{equation}

The curvature associated with the damaged undeformed manifold is of particular relevance and hence is determined in section \ref{Surface Energy}.

\section{Geometry inspired free energy for damaged solid}\label{freeenergy}

\subsection{Free Energy of an undamaged hyperelastic solid} \label{Differ_dam_FreeEnergy}

For compressible solids, the free energy takes on the following form,
\begin{equation}\label{Differ_dam_18}\begin{split}
     \psi = \hat\psi (\mathbf{C},\theta) = & k_B n N \theta\Bigg[\overline{\Lambda}_{r}\mathscr{L}^{-1}(\overline{\Lambda}_{r}) + \ln{ \frac{\mathscr{L}^{-1}(\overline{\Lambda}_{r})}{\sinh(\mathscr{L}^{-1}(\overline{\Lambda}_{r}))}}\Bigg.\\&\Bigg. - f(\overline{\Lambda}^r_{ref})\Bigg]
     + \frac{K}{2} (J-J_{ref})^2
     \end{split}
\end{equation}
where, $\overline{\Lambda}_{r} = \frac{\overline{\Lambda}}{N}$, $\overline{\Lambda} = J^{- \frac{1}{3}} \frac{1}{3}tr(\mathbf{C})^{1/2}$,  $J=\frac{\sqrt{det{g}}}{\sqrt{det{G}}}det{\mathbf{F}}$, $J_{ref} = 1$ and $\overline{\Lambda}_{rref}$ is deviatoric reference strain.

For smaller strains, the cumbersome inverse-Langevin form can be replaced with the Neo-Hookean form and the final energy can be of the form,

\begin{equation}\label{Differ_dam_18A}\begin{split}
     \psi = \hat\psi (\mathbf{C},\theta) = & k_B n N \theta\Big(\overline{\Lambda}_{r}^2 - \overline{\Lambda}^2_{rref}\Big)
     + \frac{K}{2} (J-J_{ref})^2
     \end{split}
\end{equation}

\subsection{Modification in strain energy}
Thus, for compressible solids, the strain energy due to the conformal stretching of polymer chains is specified by,
\begin{equation}\label{Differ_dam_19}\begin{split}
     \psi_e = \hat\psi_e (\mathbf{C},\theta) =& k_B n_d N \theta\Big(\overline{\Lambda}_{rdh}^2-\overline{\Lambda}_{rdh0}^2\Big) + \frac{K}{2} (J_{dh}^{0.5}-J_{dh0}^{0.5})^2
\end{split}
\end{equation}
where  $\overline{\Lambda}_{rdh} = \frac{\overline{\Lambda}_{dh}}{\sqrt{N}}$, $\overline{\Lambda}_{dh} = J_{dh}^{- \frac{1}{3}} \frac{1}{\sqrt{3}}tr(\mathbf{C}_{dh})^{1/2} = J^{- \frac{1}{3}} \frac{1}{\sqrt{3}}tr(\mathbf{C})^{1/2}$, and $J = \sqrt{det{\mathbf{F^T g F}}}$ $n_d$ is the number density of chains in the deformed damaged configuration expressed in material coordinates. Hence, $n_d = n (1-D)^3$.

Substituting the geometrically consistent stretches and Jacobian in \eqref{Differ_dam_19} we obtain the following strain energy,
\begin{equation}\begin{split}\label{Differ_dam_20A}
     \psi_e = \hat\psi_e (\mathbf{C},\theta) =& k_B n (1-D(1-h))^{3/2} N \theta\Big(\overline{\Lambda}_{r}^2-1\Big) + \frac{K}{2}(1-D(1-h))^{3/2}(J-1)^2
\end{split}\end{equation}

\subsection{Curvature Based Resistive Energy}\label{Surface Energy}

The total free energy in a damaged solid comprises of the elastic energy derived in the earlier section and a surface energy or a resistive energy which we motivate from the Ricci curvature that we determine in this section.

\begin{equation}\label{Differ_dam_21}
\hat{\psi}_{R_{dh}}(R_{dh},D) = M R_{dh}(D,\nabla D, \Delta D, H, \nabla H, \Delta H) \sqrt{det(\mathbf{G}_{dh})}
\end{equation}

where $R_{dh}(D,\nabla D, \Delta D, H, \nabla H, \Delta H)$ is the Ricci curvature, and $M$ is a material parameter quantifying the resistance of the material to the creation of a non-trivial curvature or new surfaces. $l_R$ is the width of the zone of non-trivial curvature. The Christoffel symbols associated with the Levi Civita connection for both manifolds may be derived from the compatibility condition associated with the undeformed, damaged Riemannian manifold given by,

\begin{equation}\label{Differ_dam_22}
\nabla_\alpha G_{d|{\mu\nu}} =  0
\end{equation}

The Christoffel symbols in the indicial notation are thus given by,

\begin{equation}\label{Differ_dam_23}
    \Gamma_{dMI}^L =  \frac{1}{2}G_{dh}^{KL}\left[G_{d IK,M} + G_{d KM,I} - G_{d MI,K}\right]
\end{equation}

The curvature tensor $\mathbf{R}$ can be calculated from the Christoffel symbols

\begin{equation}\label{Differ_dam_25}
\mathbf{R}^L_{dIKJ} = \partial_K \Gamma^L_{dJI} -\partial_J \Gamma^L_{dKI} + \Gamma^L_{dK A}\Gamma^A_{dJI} + \Gamma^L_{dJ A}\Gamma^A_{dKI}
\end{equation}

The components of the Ricci tensor can be calculated from the curvature tensor as follows,

\begin{equation}\label{Differ_dam_26}
\mathbf{R}^K_{dIKJ} = \partial_K \Gamma^K_{dJI} -\partial_J \Gamma^K_{dKI} + \Gamma^K_{dK A}\Gamma^A_{dJI} + \Gamma^K_{dJ A}\Gamma^A_{dKI}
\end{equation}

The Ricci scalar curvature, by definition, is a result of the contraction of the two lower indices in the Ricci tensor and is given by,

\begin{equation}\label{Differ_dam_27}
R = \mathbf{G_{dh}}^{IJ} \mathbf{R}^K_{dIKJ}
\end{equation}

With equations \eqref{Differ_dam_23}, \eqref{Differ_dam_25}, \eqref{Differ_dam_26}, and \eqref{Differ_dam_27}, the Ricci curvature results as,

\begin{equation}\begin{split}\label{Differ_dam_29}
R_{dh}(D,\nabla D, \Delta D) &= \frac{2(1-h)(1-D(1-h))\Delta D +1.5(1-h)^2\nabla D\cdot\nabla D + }{(1-D(1-h))^3}  \\&\frac{ -2D(1- D(1-h))\Delta h  + 1.5D^2 \nabla h\cdot\nabla h - (4 -D(1-h))\nabla D\cdot\nabla h}{(1-D(1-h))^3} 
\end{split}\end{equation}

The curvature is used to define the resistive energy $\psi_{R_{dh}}$ which must be positive definite and vanish when the material is fully damaged.

\begin{equation}\begin{split}\label{Differ_dam_30}
\psi_{R_{dh}} &= \left(M D^2(1-h)^2+ \frac{1}{2}M l_R^2 {det(\mathbf{G}_{dh})^{\frac{1}{3}}} R_{dh}\right) \\&= \frac{1}{\sqrt{3}}M D^2(1-h)^2 +0.75\frac{1}{\sqrt{3}}M l_R^2(1-h)^2\nabla D\cdot\nabla D +  \\& + 0.75\frac{1}{\sqrt{3}}M l_R^2D^2 \nabla h\cdot\nabla h 
\end{split}\end{equation}

The constant $M$ is the resistive modulus \cite{das2021geometrically} that quantifies the resistance offered by the material to damage, and $l_R$ is the width of the damaged area. For a case of pure damage, as described in the appendix, $M$ may be related to Griffith's fracture toughness. 

\subsection{Modified Free Energy of the Damaged Deformed Solid}
The total free energy is thus given by,
\begin{equation}\label{Total Free Energy}\begin{split}
&\psi = \psi_{R_{dh}} + \psi_e =  \frac{1}{\sqrt{3}}M D^2(1-h)^2  +0.75\frac{1}{\sqrt{3}}M l_R^2(1-h)^2\nabla D\cdot\nabla D +  \\&  + 0.75\frac{1}{\sqrt{3}}M l_R^2D^2 \nabla h\cdot\nabla h + k_B n (1-D(1-h))^{3/2} N \theta\Big(\overline{\Lambda}_{r}^2-1\Big) \\& +\frac{K}{2}(1-D(1-h))^{3/2}(J^{0.5}-1)^2
\end{split}
\end{equation}

\section{Macroforce and microforce balances}\label{Macroforce and Microforce balance}
We use the principle of virtual power to derive the macroforce and microforce balance laws. The external power is defined as,
\begin{equation}\label{Differ_dam_33}
\delta W_{ext} = \int_{\partial {V}_0} (\mathbf{t}\cdot \delta{\dot{\phi}} + \chi \delta{\dot{D}} + \gamma \delta{\dot{h}}) d A_0 + \int _{V_0} \mathbf{b} \cdot \delta{\dot{\phi}} d V_0 \end{equation}
and the internal power is given by,
\begin{equation}\label{Differ_dam_34}
\delta W_{int} = \int_{{V}_0} \mathbf{P} : \delta {\nabla_X \dot{\phi}} + \pi \delta {\dot{D}} +\beta \delta {\dot{h}} + \mathbf{\xi} \cdot \delta {\nabla{\dot{D}}}+ \zeta \delta {\Delta{\dot{h}}} + \omega \cdot \delta {\Delta{\dot{D}}} + \mathbf{\kappa} \delta{\nabla\dot{ h}} d V_0 \end{equation}
where $\pi$ and $\xi$ are the driving forces dual to $D$ and $\nabla{D}$, respectively and $\gamma$ and $\kappa$ are the driving forces dual to $h$ and $\nabla{h}$, respectively.

Equating the external and internal virtual powers and for arbitrary $\delta {\dot{\phi}}$, $\delta {\dot{D}}$ and $\delta {\dot{h}}$, the following balance laws and the boundary conditions result. First, the macro force balance is given by,
\begin{equation}\label{Differ_dam_35}
\nabla \cdot \mathbf{P} + \mathbf{b} = 0 
\end{equation}
The microforce balances for $D$ and $h$ take the form,
\begin{equation}\label{Differ_dam_36}
\nabla \cdot \mathbf{\xi} - \pi + \nabla \cdot \nabla \omega = 0 
\end{equation}
\begin{equation}\label{Differ_dam_36A}
\nabla \cdot \mathbf{\kappa} - \beta + \nabla \cdot \nabla \zeta = 0 \end{equation}
The boundary conditions are given by,
\begin{equation}\begin{split}\label{Differ_dam_37}
\mathbf{\xi}\cdot \mathbf{n}_0 = 0\quad\text{at}\quad \partial V_0 =0 \\
\mathbf{\kappa}\cdot \mathbf{n}_0 = 0\quad\text{at}\quad \partial V_0 =0 \\
\mathbf{P} \mathbf{n}_0 = \mathbf{t}\quad\text{at}\quad \partial V_0 =0 \\
\zeta \mathbf{n}_0 \cdot \mathbf{n}_l = 0\quad\text{at}\quad \partial^2 V_0 =0 \\
\omega \mathbf{n}_0 \cdot \mathbf{n}_l = 0\quad\text{at}\quad \partial^2 V_0 =0
\end{split}
\end{equation}

\section{Constitutive theory}\label{Constitutive theory}
The first and the second laws of thermodynamics may be stated as follows:
\begin{equation}\label{Differ_dam_38}
\dot{\psi} + \eta \dot{\theta} + \dot{\eta} \theta = \mathbf{P} : \dot{\mathbf{F}} + \pi \dot{D} + \beta \dot{h} + \mathbf{\xi} \cdot \nabla {\dot{D}} + \kappa \cdot \nabla {\dot{h}} + \omega \cdot \Delta {\dot{D}} + \zeta \cdot \Delta {\dot{h}} - \nabla \cdot \mathbf{q}
\end{equation}
\begin{equation}\label{Differ_dam_39}
\dot{\eta} + \frac{\nabla \cdot \mathbf{q}}{\theta} - \frac{\mathbf{q} \cdot \nabla \theta}{\theta^2} \ge 0
\end{equation}
Substituting the expression for $\nabla \cdot \mathbf{q}$ from Eqn. (\ref{Differ_dam_38}) in Eqn. (\ref{Differ_dam_39}), we obtain,
\begin{equation}\label{Differ_dam_39A}
\frac{1}{\theta}\big(\mathbf{P} : \dot{\mathbf{F}} + \pi \dot{D} + \beta \dot{h} + \mathbf{\xi} \cdot \nabla {\dot{D}} + \kappa \cdot \nabla {\dot{h}}+ \omega \cdot \Delta {\dot{D}} + \zeta \cdot \Delta {\dot{h}} -\dot{\psi} - \eta \dot{\theta} \big) - \frac{\mathbf{q} \cdot \nabla \theta}{\theta^2} \ge 0
\end{equation}

While $\mathbf{P}$,$\zeta$, $\omega$ $\xi$ and $\beta$ are all energetic $\pi$ and $\kappa$ are assumed to have both energetic and dissipative components; that is,  $\pi = \pi_{en} + \eta_D \dot{D}$, and $\beta = \beta_{en} - \eta_h \dot{h}$ where $\eta_D$ and $\eta_h$ are inverse of timescales associated with the evolution of damage and healing respectively. The constitutive form for the dissipative component ensures non-violation of the second law. The free energy and the thermodynamic forces are assumed to have the following constitutive dependencies: 
\begin{equation}\label{Differ_dam_39B}\begin{split}
&\psi = \hat{\psi}(\mathbf{F}, D, \nabla{D}, h, \nabla{h}, \theta) \text{ , } \mathbf{P} = \hat{\mathbf{P}} (\mathbf{F}, D, h,\theta)\text{ , } \xi = \hat{\xi} (\nabla{D},h)\text{ , }\\& \pi = \hat{\pi} (h,D)\text{ , }\kappa = \hat{\kappa} (\nabla{h},D)\text{ , } \beta = \hat{\beta} (h,D)\text{ , } \omega = \hat{\omega} (\Delta{D},h)\text{ , } \zeta = \hat{\zeta} (\Delta{h},h)
\end{split}\end{equation}
Using these relations, the second law may be rewritten as,
\begin{equation}\begin{split}\label{Differ_dam_39C}
&\big(\mathbf{P} - \frac{\partial{\psi}}{\partial{\mathbf{F}}}\big) : \dot{\mathbf{F}} + \big(\pi_{en}- \frac{\partial{\psi}}{\partial{D}}\big) \dot{D} + \big(\beta_{en}- \frac{\partial{\psi}}{\partial{h}}\big) \dot{h} + \big(\mathbf{\xi}- \frac{\partial{\psi}}{\partial{\nabla{h}}}\big) \cdot \nabla {\dot{h}}\\
&+ \big(\kappa- \frac{\partial{\psi}}{\partial{\nabla{D}}}\big) \dot {\nabla D}+ \big(\zeta- \frac{\partial{\psi}}{\partial{\Delta{h}}}\big) \cdot \Delta{\dot{h}}+ \big(\omega- \frac{\partial{\psi}}{\partial{\Delta{D}}}\big) \dot {\Delta D}  - \big(\eta+ \frac{\partial{\psi}}{\partial{\theta}}\big) \dot{\theta} \\
&+ \eta_D (\dot{D})^2 + \eta_h (\dot{h})^2 - \frac{\mathbf{q} \cdot \nabla \theta}{\theta} \ge 0
\end{split}\end{equation}

The above statement of the second law gives rise to the following constitutive restrictions:
\begin{equation}\label{Differ_dam_40}
    \mathbf{P} = \frac{\partial{\psi}}{\partial{\mathbf{F}}} ; \mathbf{\xi} = \frac{\partial{\psi}}{\partial{\nabla D}} ; \kappa = \frac{\partial{\psi}}{\partial{\nabla h}}; \zeta = \frac{\partial{\psi}}{\partial{\Delta h}} ; \omega = \frac{\partial{\psi}}{\partial{\Delta D}} ; \eta = - \frac{\partial{\psi}}{\partial{\theta}} ; \pi_{en} = \frac{\partial{\psi}}{\partial{D}} ; \beta_{en} = \frac{\partial{\psi}}{\partial{h}} 
\end{equation}
The heat flux assumes a Fourier form $\mathbf{q} = k_c \nabla{\theta}$, where $k_c$ is a Fourier constant.

The constitutive forms for $\xi, \beta \kappa$ and $\pi$ may be implemented in the microforce balances to yield the evolution equations for damage and healing. 
\begin{equation}\label{Differ_dam_41}
    \nabla \cdot \frac{\partial{\psi}}{\partial{\nabla D}} - \frac{\partial{\psi}}{\partial{D}}-\eta_D \dot{D} + \nabla \cdot \nabla\frac{\partial{\psi}}{\partial{\Delta D}} =0
\end{equation}
\begin{equation}\label{Differ_dam_41A}
    \nabla \cdot \frac{\partial{\psi}}{\partial{\nabla h}} - \frac{\partial{\psi}}{\partial{h}}-\eta_h \dot{h} + \nabla \cdot \nabla\frac{\partial{\psi}}{\partial{\Delta h}}=0
\end{equation}
The explicit forms for $\mathbf{P}, \xi, \kappa$ and $\pi$ are given as follows:
\begin{equation}\label{Differ_dam_42}\begin{split}
\mathbf{P} = &\mu (1-D(1-h))^{3/2} J^{-2/3}\left[\mathbf{F} - \frac{tr(\mathbf{F}^T\mathbf{F})}{3}\mathbf{F}^{-T}\right]\\
&+  \frac{K}{4} (1-D(1-h))^{3/2} (J^{0.5}-1) (1+ \tanh{p(J-1)} )J^{0.5} \mathbf{F}^{-T} \\
&+ \frac{K}{4} (1-D(1-h))^{3/2} p (J^{0.5}-1)^2 (1- \tanh{^2p(J-1)}) J \mathbf{F}^{-T}
\end{split}\end{equation}

\begin{equation}\label{Differ_dam_43}\begin{split}
\xi = 1.5\frac{1}{\sqrt{3}}M l_R^2 (1-h)^2 \nabla D 
\end{split}\end{equation}
\begin{equation}\label{Differ_dam_44}\begin{split}
\kappa = 1.5\frac{1}{\sqrt{3}}M l_R^2 D^2 \nabla h 
\end{split}\end{equation}
% \begin{equation}\label{Differ_dam_44ze}\begin{split}
% \zeta = -Ml_R^2(1-D(1-h))D
% \end{split}\end{equation}
% \begin{equation}\label{Differ_dam_44ze}\begin{split}
% \omega = Ml_R^2(1-D(1-h))(1-h)
% \end{split}\end{equation}
\begin{equation}\label{Differ_dam_45}\begin{split}
\pi_{en} &= -3/2\mu (1-D(1-h))^{1/2}(1-h)\Big(\overline{\Lambda}_{r}^2-1\Big) 
\\&- \frac{3K}{4}(1-D(1-h))^{2/3}(1-h)(J^{0.5}-1)^2 +1.5\frac{1}{\sqrt{3}}M l_R^2 D \nabla h\cdot\nabla h\\& +2\frac{1}{\sqrt{3}}{9}MD(1-h)^2
\end{split}\end{equation}
\begin{equation}\label{Differ_dam_44a}\begin{split}
\beta_{en} &= 3/2\mu (1-D(1-h))^{1/2}D\Big(\overline{\Lambda}_{r}^2-1\Big) 
\\&+ \frac{3K}{4}(1-D(1-h))^{1/2}D(J^{0.5}-1)^2 -1.5\frac{1}{\sqrt{3}}M l_R^2 (1-h) \nabla D\cdot\nabla D \\&-2 M\frac{1}{\sqrt{3}} D^2(1-h)
\end{split}\end{equation}

The microforce balance results in,
\begin{equation}\begin{split}\label{Differ_dam_46}
    &\frac{3}{2}\frac{1}{\sqrt{3}}M l_R^2 (1-h)^2 \Delta D -3 \frac{1}{\sqrt{3}}M l_R^2(1-h)D 
 \nabla h\cdot\nabla D -1.5\frac{1}{\sqrt{3}}M l_R^2D 
 \nabla h\cdot\nabla h  \\& +3/2\mu (1-D(1-h))^{1/2}(1-h)\Big(\overline{\Lambda}_{r}^2-1\Big) 
+ \frac{3K}{4}(1-D(1-h))^{1/2}(1-h)(J^{0.5}-1)^2 \\&-2\frac{1}{\sqrt{3}}MD(1+h(h-2)) -\eta_D \dot{D} =0
\end{split}\end{equation}

\begin{equation}\begin{split}\label{Differ_dam_46A}
    &2\frac{1}{\sqrt{3}}MD^2(1-h) + 1.5M\frac{1}{\sqrt{3}}l_R^2(1-h)\nabla D\cdot\nabla D+1.5M\frac{1}{\sqrt{3}}l_R^2 D^2 \Delta h-\eta_h \dot{h} \\& -3/2\mu (1-D(1-h))^{1/2}D\Big(\overline{\Lambda}_{r}^2-1\Big) 
- \frac{3K}{4}(1-D(1-h))^{1/2}D(J^{0.5}-1)^2=0
\end{split}\end{equation}

The microforce balance laws \eqref{Differ_dam_46} and \eqref{Differ_dam_46A} are used to define the evolution equations for $D$ and $h$. 

\subsection{Damage and healing evolution laws}

The evolution equations for $h$ and $D$ for pure damage and healing by the two mechanisms – diffusion and rebonding  are derived from the microforce balance laws \eqref{Differ_dam_46} and \eqref{Differ_dam_46A}. The healing variable $h$ measures the ratio of the damaged area retrieved by chain diffusion, while $D$ measures the damaged area at any instant. When the material heals by rebonding, $D$ is assumed to spontaneously reverse, and the area retrieval is measured by the scaling factor $(1-D)$.
We also model healing in which diffusion precludes rebonding. Both $D$ and $h$ are assumed to evolve, and the strong coupling between the two variables enables us to model realistically the simultaneous evolution of the healing variable for diffusion and the reversal of the damage variable for rebonding. Rate-dependent healing and damage can be modelled through the time derivatives of the mobility constants $\eta_D$ and $\eta_h$. Also, both healing and damage can be stress-driven.   

\subsubsection{Rate dependent damage model}

We retrieve the damage model of \cite{das2021geometrically} from \eqref{Differ_dam_46} when there is only damage but no healing with either of the mechanisms. The evolution of the damage variable results as, 

\begin{equation}\begin{split}\label{Differ_dam_46puredamage}
    &\frac{3}{2}\frac{1}{\sqrt{3}}M l_R^2 \Delta D +3/2\mu (1-D)^{1/2}\Big(\overline{\Lambda}_{r}^2-1\Big) 
+ \frac{3K}{4}(1-D)^{1/2}(J^{0.5}-1)^2 \\&-2\frac{1}{\sqrt{3}}MD-\eta_D \dot{D} =0
\end{split}\end{equation} 

This relation must satisfy the $\Gamma-convergence$ criterion \cite{borden2012phase, das2021geometrically} from which the relation between $M$ and $G_c$ ($G_c = M l_R$)  is established. Detailed calculations have been shown in the appendix.

\subsubsection{Intrinsic healing by diffusion}

Healing by diffusion is a spontaneous intrinsic, non-stress-driven process where chains diffuse into the damaged region, healing it with a distinct mass with properties different from the virgin material. Healing by diffusion recovers the effective stress-bearing area but the existing damage is not reversed. 

Incorporating, these features, we modify \eqref{Differ_dam_46A}

\begin{equation}\begin{split}\label{Differ_dam_46Ahealdiffusion}
    &2\frac{1}{\sqrt{3}}D^2(1-h) + 1.5\frac{1}{\sqrt{3}}l_R^2(1-h)\nabla D\cdot\nabla D+1.5\frac{1}{\sqrt{3}}l_R^2 D^2 \Delta h-\tau_h \dot{h} =0
\end{split}\end{equation}

 Further we have scaled the equation by $\frac{1}{M}$, and introduced the chain-diffusion healing timescale, $\tau_h = \frac{\eta_h}{M}$. Several researchers have found that this timescale is similar to the stress-relaxation timescale, which controls material creep.

\subsubsection{Rebonding of reversible bonds} 

The rebonding healing mechanism is a spontaneous damage reversal phenomenon that is derived from \eqref{Differ_dam_46} by removing stress-based driving terms. 

\begin{equation}\begin{split}\label{Differ_dam_46reversaldamage}
    &-2\frac{1}{\sqrt{3}}D+\frac{3}{2}\frac{1}{\sqrt{3}} l_R^2 \Delta D  -\tau_D \dot{D} =0
\end{split}\end{equation}

Similar to \eqref{Differ_dam_46Ahealdiffusion}, the equation is scaled by a factor of $\frac{1}{M}$, to introduce the rebonding timescale $\tau_D = \frac{\eta_D}{M}$.

\subsubsection{Rebonding of reversible bonds preceded by inter-chain diffusion}
Where diffusion of chains and rebonding follow each other and the rebonding process is dependent on the diffusion of chains and vice-versa, both area recovery through $h$ and reversal of damage through $D$ occurs. 

\begin{equation}\begin{split}\label{Differ_dam_46coupleddamage}
    &\frac{3}{2}\frac{1}{\sqrt{3}}l_R^2 (1-h)^2 \Delta D -3 \frac{1}{\sqrt{3}} l_R^2(1-h)D 
 \nabla h\cdot\nabla D -1.5\frac{1}{\sqrt{3}}l_R^2D 
 \nabla h\cdot\nabla h   -\tau_D \dot{D} =0
\end{split}\end{equation}

\begin{equation}\begin{split}\label{Differ_dam_46Acoupledheal}
    &2\frac{1}{\sqrt{3}}D^2(1-h) + 1.5\frac{1}{\sqrt{3}}l_R^2(1-h)\nabla D\cdot\nabla D+1.5\frac{1}{\sqrt{3}}l_R^2 D^2 \Delta h-\tau_h \dot{h} =0
\end{split}\end{equation}

need to be solved simultaneously.

\section{Numerical simulations}\label{miscellaneoussimulations}

We implement the damage and the healing models in a set of numerical simulations to observe their ability to model some key features such as rate-dependent non-local damage, non-linear unloading during healing, recovery of stiffness and fracture toughness, increase of recovery with healing time and timescales of diffusion and recovery, influence on damage-width on healing rate etc. The first set of simulations reproduces the numerical experiments of \cite{darabi2012continuum} in an incompressible elastomer in a homogeneous stress state. This is followed by simulations of the diffusion-based healing of damage of various initial widths in a one-dimensional bar to determine the effect of width on the extent of healing. Finally, the model is used to predict fracture in virgin and healed supramolecular elastomers and understand post-healing fracture properties of these materials as observed by \cite{cordier2008self}.

\subsection{Coupled rate-dependent damage and healing through homogeneous simulations}

A typical damage-healing model for general polymers must possess certain basic features such as stiffness-recovery, non-linear unloading under damage reversal \cite{ortiz1985constitutive} that can be established through single material point simulations of the numerical experiments of \cite{darabi2012continuum}. These experiments, comprising repeated cycles of loading and healing in a polymer, investigate the damage response of healed and virgin materials. We also use these simulations to carry out parametric investigations.

We subject, an incompressible elastomer to a loading-unloading-reloading cycle 
 in which quasistatic tensile loading is applied followed by a period of combined healing followed by reloading.  

Equations \eqref{Differ_dam_46puredamage}, \eqref{Differ_dam_46Ahealdiffusion}, \eqref{Differ_dam_46coupleddamage} are solved. The incompressibility of the elastomer is imposed through deformation gradient of the form, 

\begin{equation}\mathbf{F} = \begin{bmatrix} \lambda & 0 & 0 \\ 0 & \lambda & 0 \\ 0 & 0 & \lambda \end{bmatrix}\end{equation}

in the equation \eqref{Differ_dam_42}, with $\lambda$ being the stretch.

The simulations yield results that qualitatively compare well with \cite{darabi2012continuum}. The stiffness and the fracture toughness increase with both the healing time and the timescale $\tau_h$.
 
\begin{figure}
\centering
    \begin{subfigure}[b]{0.8\textwidth}{\includegraphics[width=\textwidth]{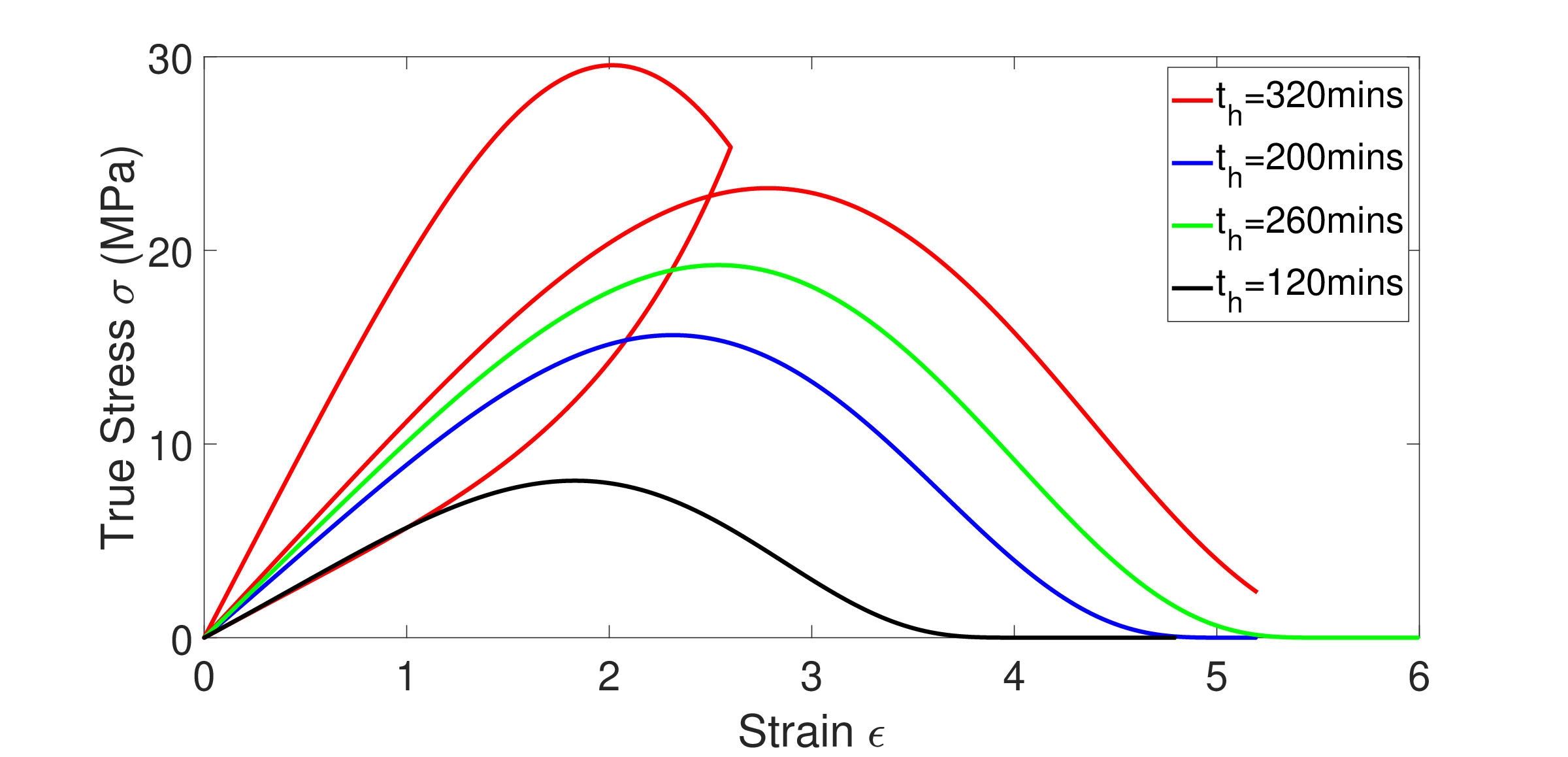}} 
    \caption{}
    \end{subfigure}
    \begin{subfigure}[b]{0.8\textwidth}{\includegraphics[width=\textwidth]{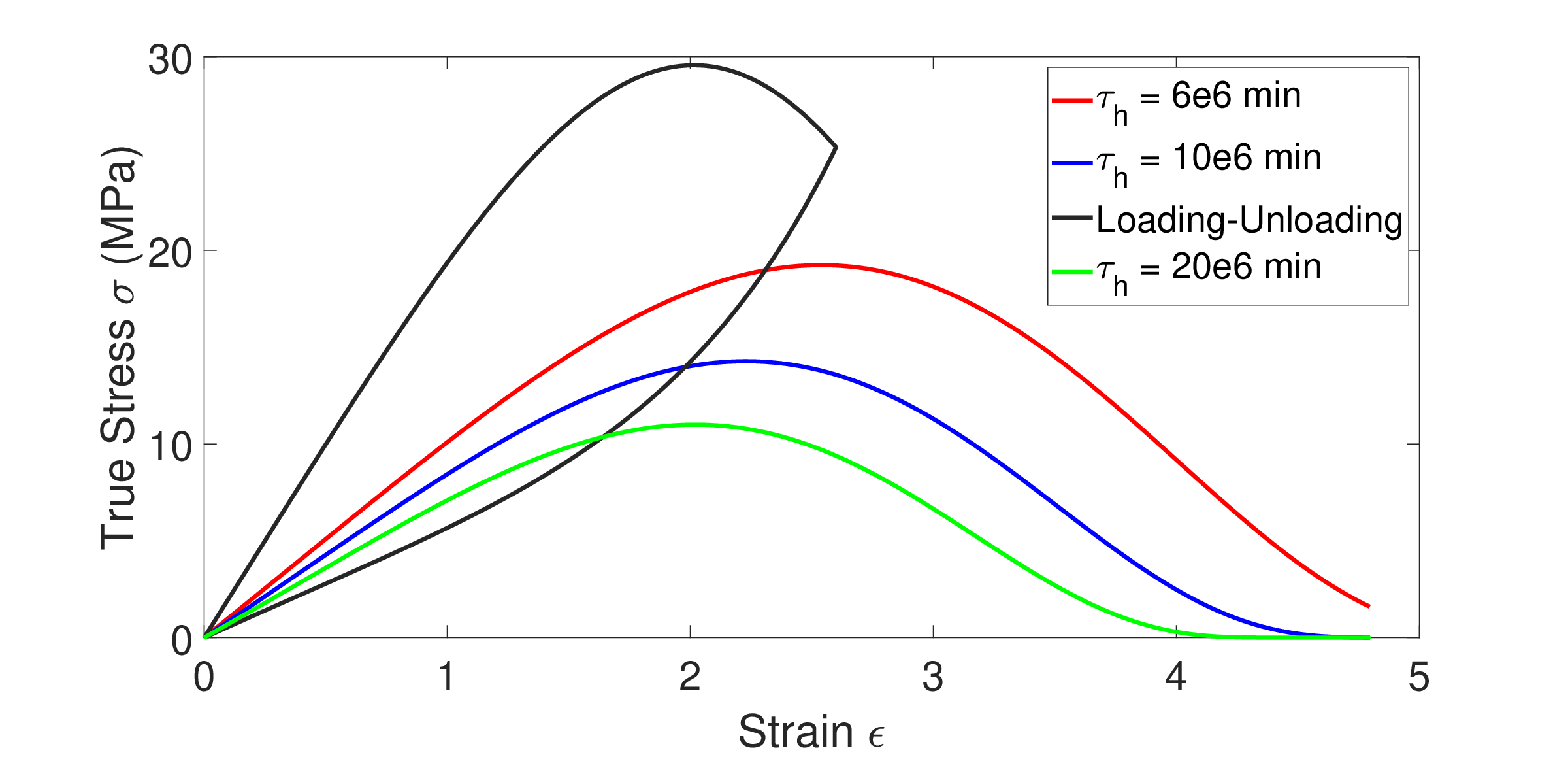}} \caption{}
    \end{subfigure}

\caption{\label{HomogeneousSelfhealetah} Fig (a): Stress vs strain response in an incompressible elastomer under loading-unloading and healing-reloading experiment \cite{darabi2012continuum} Post loading and unloading material was then allowed to heal for $t_h$ healing time before being reloaded again. The peak stress post-healing increases with an increase in $t_h$ indicating a gradual recovery of the fracture toughness. Nonlinear stress vs strain relation during unloading due to damage reversal in line with \cite{darabi2012continuum,ortiz1985constitutive} Fig (b): Stress-strain curves for various healing timescales $\tau_h$. As $\tau_h$ increases, healing occurs at a slower rate and stiffness and fracture toughness recovery decreases.}
\end{figure}

\subsection{1D simulation: Influence of width of interfacial damage on the extent of healing}

In this section, the model is used to predict the influence of the damage width on the healing rate in a bar with a centrally placed damage zone. It is used to qualitatively predict the observations of \cite{amaral2017stimuli} that in the case of diffusion-entanglement-driven healing, narrower damage heals faster than wider damage. This is because narrow damage requires a smaller distance through which chains must travel to entangle. 

The equation \eqref{Differ_dam_46Ahealdiffusion} is solved in one-dimension using the finite difference method. The one-dimensional form for the equation is given as follows,

\begin{equation}\begin{split}\label{Differ_dam_46AHHIntrinsichealoneD}
    2\frac{1}{\sqrt{3}}D^2(1-h) + 1.5\frac{1}{\sqrt{3}}l_R^2(1-h)\left(\frac{\partial{D}}{\partial{X}}\right)^2+1.5\frac{1}{\sqrt{3}}l_R^2 D^2 \frac{\partial^2 h}{\partial X^2}-\tau_h \dot{h}=0
\end{split}\end{equation}

\begin{figure}
\centering
\includegraphics[width=0.8\textwidth]{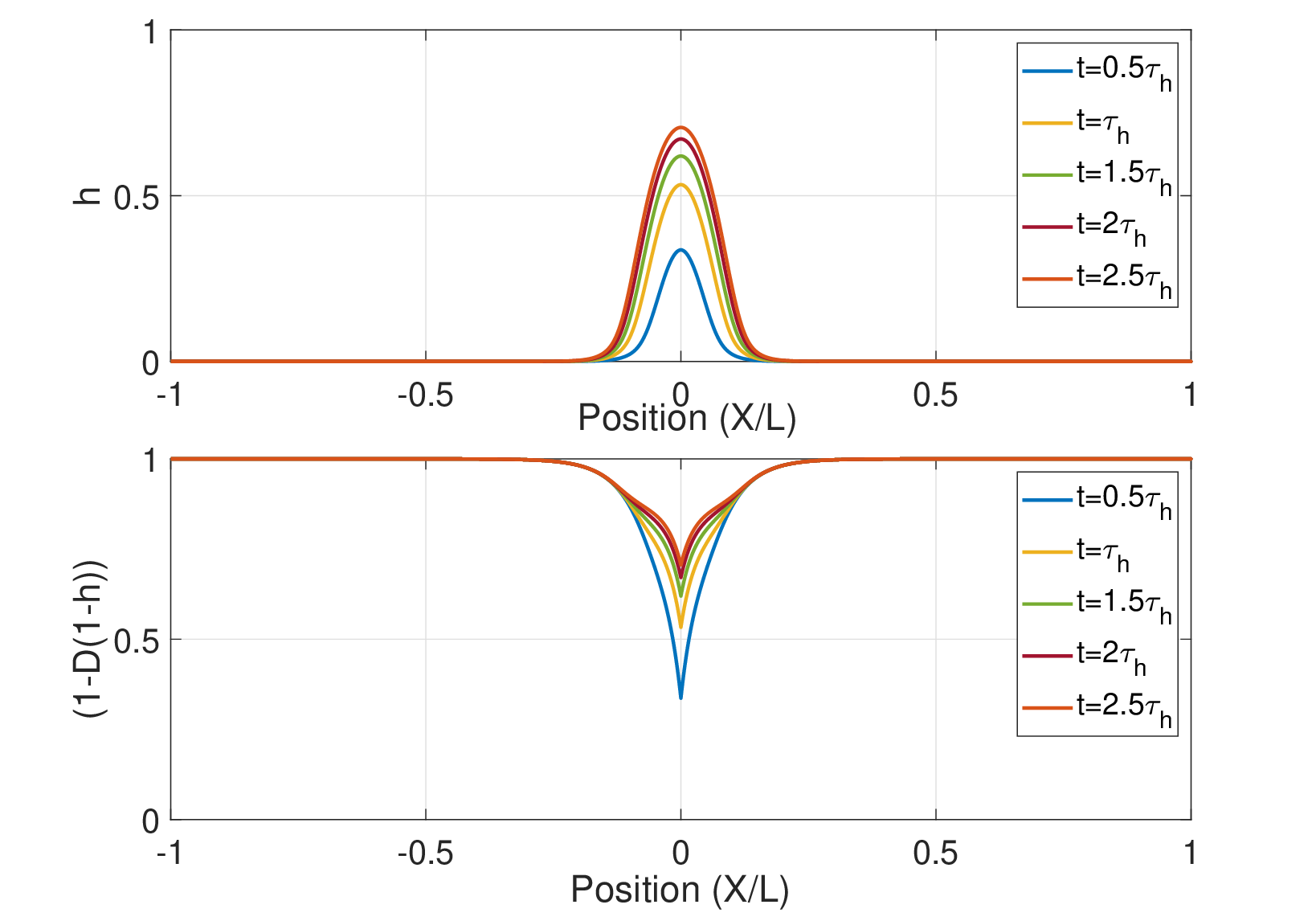}\centering
        \caption{Time snapshots of healing $h$ vs $X$ and degradation $(1-D(1-h))$ vx $X$ for healing of a crack ($D_0=1$) of width $L_{R0}$ of 0.05$L$ with a $\tau_h$ of 40 mins and a healing width $l_R$ of 0.1$L$. The intensity of healing increases and that of the degradation decreases with time.}\label{onedimensionspatialhealdegradation}
    \end{figure}

\begin{figure}
\centering
\includegraphics[width=0.8\textwidth]{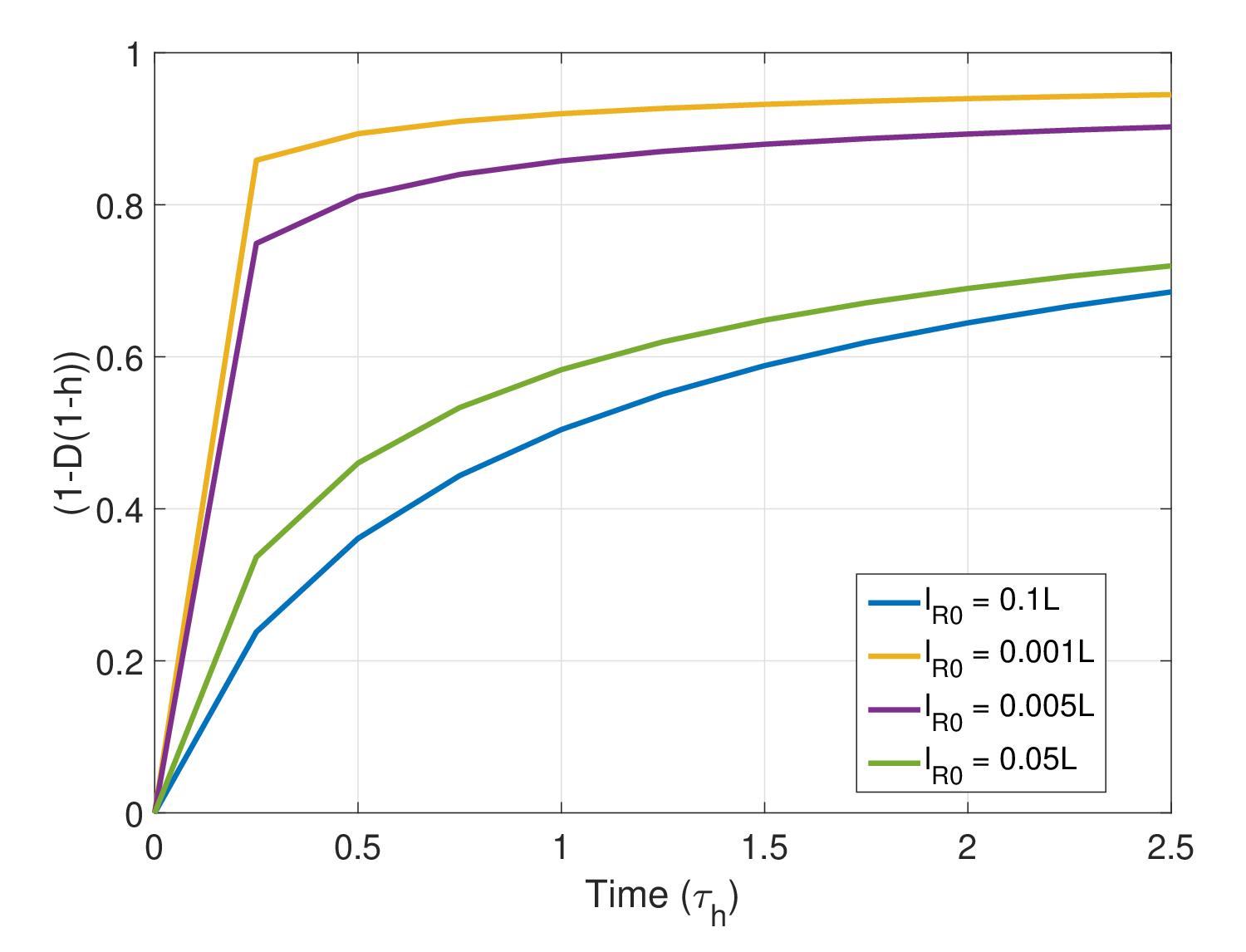}\centering
        \caption{$(1-D(1-h))$ at $X=0$ with time for different $L_{R0}$ (0.002 L, 0.05 L, 0.1 L). With the decrease in $L_{R0}$ i.e. for narrower damage, the recovery of properties as measured by the degradation function is faster}\label{onedimensiontimedegradation}
    \end{figure}

The centrally placed diffused damage is modelled by specifying the damage variable as, $D(X,t=0) = D_0 \exp{-\frac{X}{l_{R0}}}$.  At $X=0$, the center of the rod, $D$ is maximum with a value of $D_0$ and $l_{R0}$ is the width of the damage. $D_0$ for a fully developed crack is 1 and less than 1 otherwise. $h$ is initially 0 everywhere and must satisfy the following boundary conditions - $\frac{\partial{h}}{\partial{X}} = 0$ at $X=\infty$ and $X=0$. Figure \ref{onedimensionspatialhealdegradation} shows the evolution of $h$ and $1-D(1-h)$, the stiffness degradation function during the healing of a crack of width $l_{R0}=0.05L$ duration $t_h$. With time healing increases, while the crack width remains constant.  Figure \ref{onedimensiontimedegradation} shows the variation of the degradation function with time for different initial $l_{R0}$. Inline with \cite{amaral2017stimuli}, recovery is faster for a smaller $l_{R0}$.

\subsection{Intrinsic healing of an interfacial crack in a supraelastomer}

The model is now used to investigate the pre-healed and post-healed fracture properties of a supermolecular elastomer and to reproduce the experiments of \cite{cordier2008self}. \cite{cordier2008self} observed that these supramolecular elastomers heal at room temperature within observational timescales. The primary healing mechanism is observed to be the rebonding of non-equilibrated hydrogen bonds, and it is not preceded or accompanied by chain diffusion, as the relaxation timescale is in the order of weeks \cite{cordier2008self} and much larger than the observational timescale of interest \cite{grande2015interfacial} which is of the order of a few hours. 

In the experiment, a dogbone specimen is subject to a quasistatic uniaxial tension until fracture. Following that, the fractured pieces are brought in contact and allowed to heal at room temperature for different healing times $t_h$ - 15mins, 30mins, 60 mins, 120 mins and 180 mins. The healing process is followed by the reloading of the specimen up to fracture. The experimental stress-strain curves are shown in dotted lines in \ref{Cordiersimulationexpt}. The elastic response of the healed specimen is independent of $t_h$. The strain to fracture, however, increases with healing time. Through our simulations, we model this dependence of strain on $t_h$, reaching to some conclusions on the recovery of mechanical properties $t_h$.

The simulation of the experiment comprises the following steps - 
 
\begin{itemize}
\item Stress-driven damage and fracture of the virgin specimen 
\item Rebonding or reversal of damage in the fractured specimen. A chain diffusion or the stress relaxation timescale $\tau_h$ is much higher than $\tau_D$ \cite{cordier2008self} compared to our observational timescale, thus healing can be simulated by damage reversal alone.
\item Stress-driven damage and fracture of the healed specimen with recovered material properties. 
\end{itemize}

The complete superposition of the elastic parts of the stress-strain curves of the material with various healing times implies that healing only affects the fracture toughness, hence the resistivity modulus. Our numerical experiment first simulates the fracture of healed specimens to determine the relationship between the healing time and the resistivity modulus, and then through the healing model, which in this case is the reversal of damage, the rebonding timescale, resistivity modulus of the virgin material is determined.

\subsubsection{Fracture in healed specimen}

The linear momentum balance \eqref{Differ_dam_35} with the constitutive equation for stress \eqref{Differ_dam_42} and the rate-dependent damage evolution $D$, \eqref{Differ_dam_46puredamage} are solved with a 2D finite element code using FeNICS open-source packages. The geometry comprises a rectangular domain of size 50mm X 13mm X 2mm representing the gage length of the dogbone specimen. It is meshed using linear quadratic and triangular elements with maximum and minimum sizes of mm and mm, respectively. A displacement on the top boundary is applied at a constant strain rate of 0.0004/sec. Numerical stability issues arising due to the rapid damage evolution characteristic of brutal damage are solved by linearisation of terms \cite{ang2022stabilized} and small load step size. However, the simulation was terminated before the crack could completely propagate as the strain to fracture does not significantly change post-peak and the computational time required for each of the complete simulations is prohibitively large.

The following material constants in the equations are determined from the curve corresponding to a healing time of 180 mins -  $\mu = 2.7276\times 10^{-3} MPa$, $l_R = 0.0962mm$, $\kappa = 4.50054\times 10^{-3} MPa$. The recovered modulus of resistivity, $M_h$ which is a function of the healing time, $t_h$ must be used instead of the modulus for the virgin material. $M_h$ as a function of the healing time $t_h$ is determined from the healing model in the following section

\subsubsection{Healing by damage reversal and recovered resistive modulus}

The fracture in the dog bone specimen is a straight crack perpendicular to the longitudinal axis. The healing process appears to be uniform in width across the entire length of the crack. Assuming that the healing progresses at the same rate at all points along the crack, it is sufficient to solve the damage reversal equation \eqref{Differ_dam_46reversaldamage} along the longitudinal axis. The one-dimensional form for the equation is given by,

\begin{equation}\label{Damage reversalsupra}\begin{split}
    &\frac{3}{2}\frac{1}{\sqrt{3}} l_R^2 \frac{\partial^2 D}{\partial X^2} -2\frac{1}{\sqrt{3}}D -\tau_D \dot{D} =0
\end{split}\end{equation}

where $l_R$ represents the width of the fracture zone as well as the healed zone where the non-equilibrated hydrogen bonds are located.

Initially, a diffused damage represented by $D(X,0)=exp(-\frac{X}{\sqrt{3}l_R})$ is centered at $X=0$. Given the boundary conditions $D=0 at X=\infty$ and $\frac{\partial{D}}{\partial{X}}=0 at X=0$, the equation is solved analytically to yield the closed-form expression for $D$ as,

\begin{equation}\label{Damage reversalsupraoneD healed}\begin{split}
D(X,t) = \exp\left(-\frac{X}{\sqrt{3} l_R}\right) \exp\left(-\frac{\sqrt{3}}{2 \tau_D} t_h\right)
\end{split}\end{equation}

$D$ gives the spatial distribution of the residual damage after healing has occurred for a time period of $t_h$. When we substitute $D$ in the surface energy expression \eqref{Surface Energy}, we obtain a loss in the resistive capability of the material.

\begin{equation}\label{Differ_dam_70}\begin{split}
\int^\infty_{-\infty}\psi_{R_{dh}} dX= M l_R (\exp{-\frac{\sqrt{3}}{2 \tau_D} t_h})
\end{split}\end{equation}

The healed fracture toughness is $M l_R\left(1- (\exp{-\frac{\sqrt{3}}{2 \tau_D} t_h})\right)$ which yields the following expression for the healed modulus $M_h$, 

\begin{equation}\label{healedmodulus}
M_h = M (1-(\exp{-\frac{\sqrt{3}}{2 \tau_D} t_h}))
\end{equation}

The expression \eqref{healedmodulus} indicates that $M_h$ recovers completely at an infinite time. While this is in line with the fracture toughness experiments conducted by \cite{grande2015interfacial} where a complete recovery is never practically obtained, 
 the variation of $M_h$ with $t_h$ following a power law is not in line with \cite{grande2015interfacial} or \cite{wool1985properties}. Both studies assume chain diffusion as the primary healing mechanism, which may explain this. To obtain the variation of $M_h$ with time $t_h$, experiments must be carried out which is beyond the scope of the current article.

 The material parameters $\tau_D$, and $M$,  the rebonding timescale are obtained from the stress-strain curves corresponding to 15, 60, and 180 mins.We determine them to be 198.2 mins and 23 MPa, respectively. 

\begin{figure}[!]
\centering
\includegraphics[width=1\textwidth]{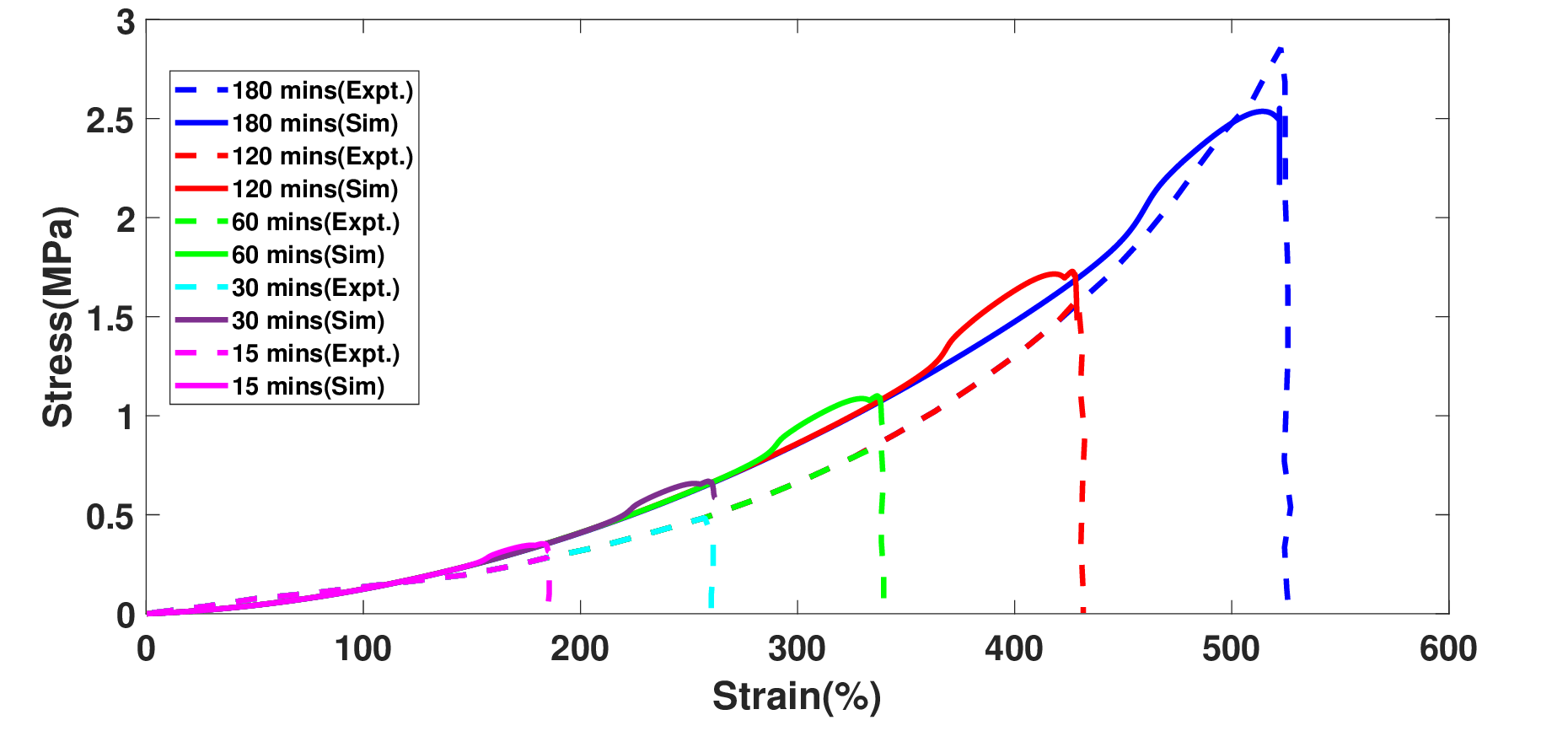}
\caption{\label{Cordiersimulationexpt} The figure shows stress-strain curves as observed in \cite{cordier2008self} and those obtained by the finite element model. The accurate prediction of strains to fracture indicate a good estimation of the modulus $M_h$. The good correlation between experiments and simulations establish our assumption that the stiffness recovers much faster and is independent of $t_h$, the healing time. An abrupt increase in stresses in the vicinity of the peak is observed due to numerical stability issues.}
\end{figure}

The $M_h$ values for 120 mins and 30 mins are obtained from the expression, \eqref{healedmodulus} and implemented in the finite element simulation for the fracture. 
Figure \ref{Cordiersimulationexpt} shows the simulation and experimental stress-strain curves for various degrees of healing. Overall, the numerically obtained curves compare well with experiments and superpose with each other in the elastic regime. The strains to fracture have been accurately predicted for $t_h =120 mins$ and $t_h = 30 mins$. There is a sudden increase of stress in the vicinity of the peak due to numerical stability issues but the peak stress correlate well with those of the experiments. 

%%%%%%%%%%%%%%%%%%%%%%%%%%%%%%%%%%%%%%%%%%%%%%%%%%%%%%%%%%%%%%%%%%%%%%
\section{Conclusions}\label{Conclusion}
We need simple damage-healing theories that are consistent with thermodynamics and can predict autonomic interfacial self-healing in elastomers. These theories can be used to design materials and for failure and healing analysis of real-life structures. Models available in the literature are complex and may fail in their predictions as they lack the necessary timescales and lengthscales. The very few phase-field-inspired approaches use surface energy prescriptions that result in damage and healing evolution equations that lack a physical basis. This article prescribes a geometry-based damage-healing theory for autonomic healing in elastomers. The theory is built on a geometry-based framework developed in \cite{das2021geometrically} where damage is postulated to induce an incompatibility in the Euclidean material manifold, transforming it into a Riemannian manifold. We can quantify the incompatibility by using a Ricci curvature that determines the resistive surface energy. We assume that healing restores the manifold to a Euclidean manifold, either through the reversal of the damage or through rescaling the damage by a healing variable that quantifies damage recovery. The microforce balances dictate the evolution of the damage and healing variables during the fracture and healing processes. The theory focuses on the following healing mechanisms: rebonding of reversible bonds, diffusion of chains across the damaged interface and rebonding preceded by chain diffusion and entanglement. The theory models rebonding through the reversal of damage, while the evolving healing variable models chain diffusion and entanglement. Both processes are intrinsic and evolve with respective timescales. Additionally, the microforce balance for the damage variable can be used to derive a rate-dependent damage model similar to \cite{das2021geometrically}. The resistive modulus of the virgin material and the healed material can be related to corresponding fracture toughnesses through $\Gamma-convergence$ criteria \cite{borden2014higher}.

The predictive capabilities of the model are explored through homogeneous, one-dimensional and two-dimensional finite element and finite difference simulations. The homogeneous simulations are those of the numerical experiments of \cite{darabi2012continuum}. In obtaining a good qualitative comparison with their results, our theory predicts a correct coupling between the healing variable and the damage variable during reversal—a feature essential to rebonding preceded by diffusion. Additionally, our theory accurately predicts non-linear stress-strain behaviours during unloading and healing, as well as the increase in peak stress with healing time. Through one-dimensional simulations of the healing of diffused damage in a bar, we show that the model predicts the influence of the width of the damage on the healing rate accurately. This is, however, limited to healing by chain diffusion and entanglement.

Finally, the model is used to predict how supramolecular elastomers heal spontaneously when cracked pieces are brought together at room temperature. The model shows that the resistive modulus and hence the fracture toughness recover exponentially with healing time during the rebonding of broken hydrogen bonds. The rate-dependent damage model with the resistive modulus-healing time relationship is validated using two-dimensional simulations of the uniaxial tension experiments of \cite{cordier2008self} using the FeNICS finite element package. The exponential recovery of fracture toughness with healing time is in line with observations of \cite{grande2015interfacial}, though it differs from those of \cite{wool1981theory} as the healing mechanism is different. 

The model is capable of a comprehensive prediction of damage and the intrinsic healing of elastomers with just a few parameters — healing and damage timescales, resistivity moduli, and a length scale of diffusion. These parameters can easily be determined through macroscopic fracture experiments similar to \cite{cordier2008self} with cyclic loading without any mesoscale investigations or rigorous curve fitting. Such simplicity makes the formulation much superior to existing prescriptions for material design or structural and failure analysis. Further, the geometric consistency of the theory gives it a solid physical base, which eliminates the necessity of ad hoc prescriptions for degradation functions and surface energy densities.

Temperature-dependent prescriptions for the timescales can introduce thermal dependencies. Furthermore, we can extend this model to other intrinsic healing mechanisms with minimal modifications. The model can also be extended to extrinsic and autogenous healing by incorporating the effect of encapsulated healing agents and other stimuli in the free energy or as separate internal variables. Additionally, by introducing irreversibility conditions through history variables, we can decouple healing from damage.

\section{Reference}
\bibliography{reference}
\bibliographystyle{elsarticle-harv} 

\section{Appendix: Pure Damage, no Self-Healing}\label{PuredamageSelf-Healing}

To derive the relationship between $M$, the resistive modulus and the fracture toughness $G_c$ we adopt the process of enforcing $\Gamma$-convergence on the surface energy density $\Psi_{Rd}$ in a one dimensional bar with a centrally positioned crack . The process has been discussed in some details in \cite{das2021geometrically} and \cite{borden2014higher}. According to the criteria, the surface energy density must converge to the fracture toughness $G_c$ in the limit of $l_R -> 0$

For that the equation \eqref{Differ_dam_46puredamage} is rewritten in a one-dimensional form, ignoring the rate dependent term, as follows,

\begin{equation}\label{Differ_dam_65App}\begin{split}
1.5\frac{1}{\sqrt{3}}M l_R^2 \frac{\partial^2 D}{\partial X^2} -2\frac{1}{\sqrt{3}}MD =0
\end{split}\end{equation}.

The boundary conditions are $D=1$ at $X=0$ and $D=0$ at $X=\infty$, 

This equation has the closed-form solution of the form,

\begin{equation}\label{Differ_dam_69App}\begin{split}
D = \exp{-\frac{2 x}{\sqrt{3}l_R}}
\end{split}
\end{equation}

Substituting the solution of $D$ in the surface energy density and integrating the surface energy over the whole domain [$-\infty, \infty$],
\begin{equation}\label{Differ_dam_70App}\begin{split}
\int^\infty_{-\infty}\psi_{R_{dh}} dX=&  \int^\infty_{-\infty}\frac{1}{\sqrt{3}}M D^2 +\frac{0.75}{\sqrt{3}}M l_R^2\left(\frac{\partial{D}}{\partial{X}}\right)^2 dX
= M l_R
\end{split}\end{equation}

Since the surface energy density must converge to $G_c$ as $l_R-> 0$, we obtain $G_c = M l_R$. 
\end{document}